\documentclass[10pt,letterpaper]{article}

\usepackage{ccn}
\usepackage{pslatex}
\usepackage{apacite}
\usepackage{hyperref}
\usepackage{natbib}
\usepackage{lineno}
\usepackage{graphicx}
\usepackage{amsmath}

\usepackage{epigraph}
\title{Sparks of cognitive flexibility: self-guided context inference for flexible stimulus-response mapping by attentional routing}
 
\author{
    {\large \bf Rowan P. Sommers\textsuperscript{†}, Sushrut Thorat\textsuperscript{†}, Daniel Anthes, Tim C. Kietzmann} \\
    \texttt{\{rsommers, sthorat, danthes, tkietzma\}@uos.de} \\
    Institute of Cognitive Science, University of Osnabrück, Germany; \textsuperscript{†}Equal contribution.
}

\begin{document}

\maketitle

\section{Abstract}
{
\bf
Flexible cognition demands discovering hidden rules to quickly adapt stimulus-response mappings. Standard neural networks struggle in such tasks requiring rapid, context-driven remapping. Recently, Hummos (2023) introduced a fast-and-slow learning algorithm to mitigate this shortcoming, but its scalability to complex, image-computable tasks was unclear. Here, we propose the Wisconsin Neural Network (WiNN), which extends Hummos' fast-and-slow learning to image-computable tasks demanding flexible rule-based behavior. WiNN employs a pretrained convolutional neural network for vision, coupled with an adjustable “context state” that guides attention to relevant features. If WiNN produces an incorrect response, it first iteratively updates its context state to refocus attention on task-relevant cues, then performs minimal parameter updates to attention and readout layers. This strategy preserves generalizable representations in the sensory and attention networks, reducing catastrophic forgetting. We evaluate WiNN on an image-based extension of the Wisconsin Card Sorting Task, revealing several markers of cognitive flexibility: (i) WiNN autonomously infers underlying rules, (ii) requires fewer examples to do so than control models reliant on large-scale parameter updates, (iii) can perform context-based rule inference solely via context-state adjustments—further enhanced by slow updates of attention and readout parameters, and (iv) generalizes to unseen compositional rules through context-state updates alone. By blending fast context inference with targeted attentional guidance, WiNN achieves “sparks” of flexibility. This approach offers a path toward context-sensitive models that retain knowledge while rapidly adapting to complex, rule-based tasks.}
\begin{quote}
\small
\textbf{Keywords:} 
cognitive flexibility; attention; routing; continual learning; compositionality; neuroconnectionism
\end{quote}

\epigraph{
  ``In times of change learners inherit the earth; 
  while the learned find themselves beautifully equipped 
  to deal with a world that no longer exists.''
}{
  --- Eric Hoffer, \emph{Reflections on the Human Condition} (1973), Aphorism 32.
}

\section{Introduction}

Adaptive, context-aware behavior is central to cognition~\citep{diamond2013executive}. A single stimulus can prompt very different responses depending on the situation: a monkey quickly grabs a piece of fruit when it is hungry and safe but ignores the same fruit if it has just eaten or senses danger. Likewise, a green light usually signals “go”—unless a traffic controller says otherwise. These examples highlight the crucial role of context in flexibly shaping stimulus-response (S-R) mappings. Understanding how the brain and machines can accomplish this flexibility remains a major challenge in the study of intelligence. This topic is especially pressing in the domain of AI, as one common criticism of artificial neural networks (ANNs) is that they are inflexible and can only slowly learn a single (albeit arbitrarily complex) S-R mapping while requiring many training examples.

\begin{figure*}[ht]
\begin{center}
\includegraphics[width=0.9\textwidth]{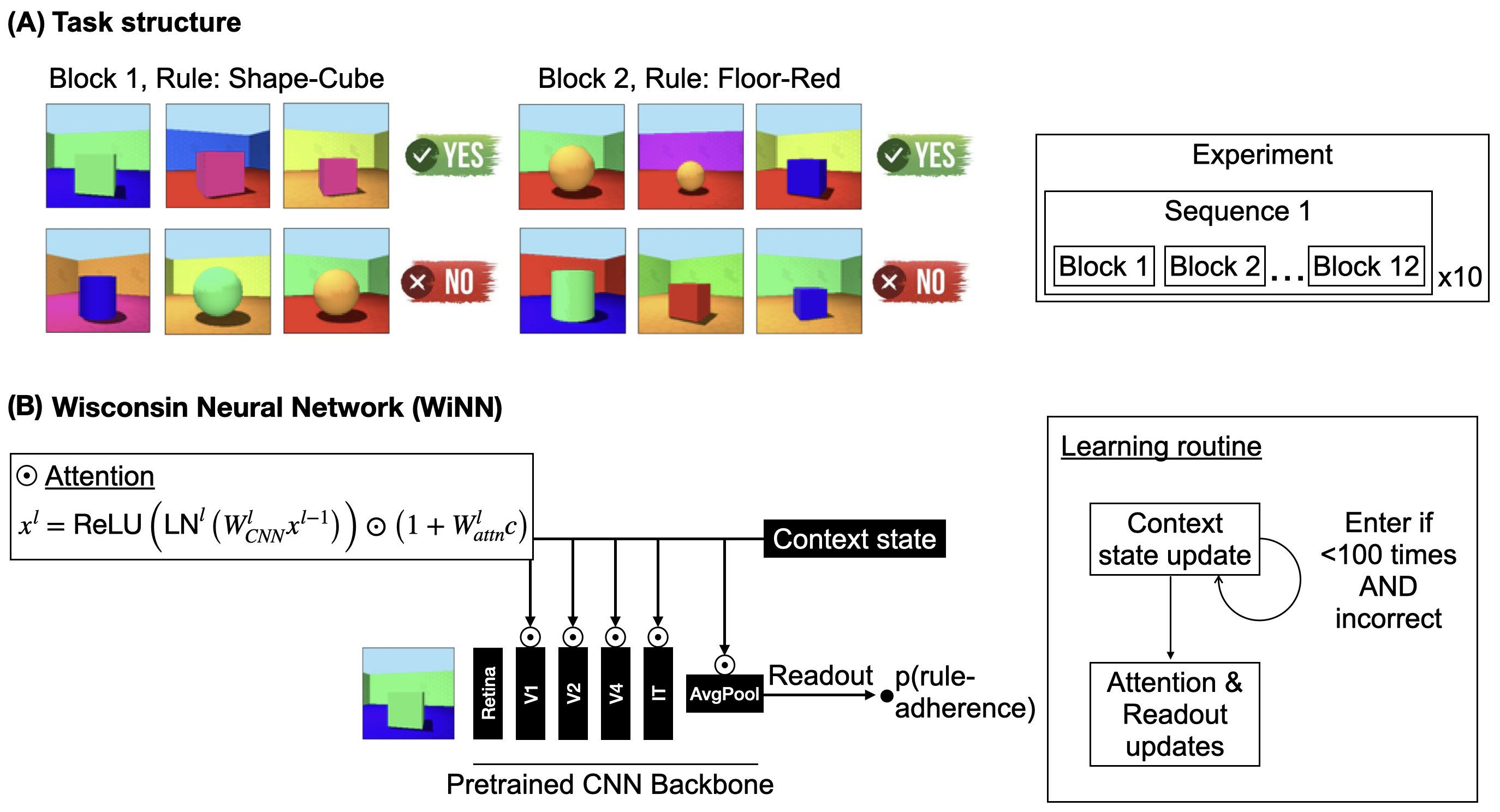}
\end{center}
\caption{\textbf{Setup.} (A) During an experiment, images are presented in succession. For each image, the task is to decide whether it adheres to a hidden rule. The hidden rule switches after $800$ images, forming a ``Block.'' 
    (B) The Wisconsin Neural Network (WiNN) is built for flexible rule inference over complex image streams. A pretrained convolutional neural network (CNN) maps the image to a response that is modulated by the inferred context. The left panel illustrates how the context state $c$ modulates the $l^\mathrm{th}$ CNN layer. Whenever the CNN produces an error, $c$ is updated iteratively to adjust the attentional weights and remap the network’s response. Subsequently, or if the response is already correct, a single update is applied to the attention and readout parameters.} 
\label{fig1}
\end{figure*}

A classic test of cognitive flexibility is the Wisconsin Card Sorting Test (WCST)~\citep{berg1948simple,grant1948behavioral}. In this task, participants sort cards according to a hidden rule — sorting by color, shape, or number — and must continuously infer and adapt to the correct rule (or “context”). Frontal lobe areas have been implicated in detecting this context~\citep{milner1963effects,koechlin2003architecture,donoso2014foundations}, while frontoparietal networks direct visual attention toward the relevant features~\citep{ungerleider2000mechanisms,corbetta2002control,boshra2022attention}. Despite extensive research, the computational details of how they interact to infer context and how they update S-R mappings remain unclear. Here, we explore multi-stage learning processes that may underlie such flexibility. As a model system, we consider their integration into artificial neural networks (ANNs, \citet{doerig2023neuroconnectionist}). 

Recent computational work by Hummos \citep{hummosthalamus} offers a promising direction for faster context adaptations in ANNs. Hummos’s framework uses a fully connected recurrent neural network (RNN) whose S-R mapping is modulated by a learned context vector. Crucially, the learning proceeds in two stages: First, when an error occurs, a sequence of gradient descent steps is applied to change the context state (i.e. context activation pattern) in order to “remap” stimuli to a correct response while using the RNN’s existing feature extractors~\citep{cheung2019superposition,thorat2019modulation}. Then, if this context-based remapping fails, the network weights themselves are adjusted. This approach not only improves flexibility and speeds up adaptation, but also protects the underlying feature extraction from drastic changes, mitigating catastrophic forgetting. However, whether this overall approach can be extended to high-dimensional, image-based tasks — such as those inspired by the WCST — remains to be seen~\citep{sandbrink2025flexible,Bellman1961}.

Here, we introduce the Wisconsin Neural Network (WiNN) to expand a fast-and-slow learning approach to more complex, visual rule-based categorization. WiNN simulates the ventral visual stream (via a convolutional neural network) and integrates a context-inference mechanism, akin to frontal cortex functions, that adapts visual inference via context-dependent attentional modulation. We show that WiNN excels on a WCST-like task, notably achieving near-perfect performance with exposure to fewer examples than control models. Moreover, once the network has sufficient exposure to the task, it can infer the active rule and select the correct response solely by updating its context state, without further changes to its weights. This fast, context-driven remapping extends to previously unseen simple and compositional rules. In this way, WiNN demonstrates what we call “sparks of cognitive flexibility,” suggesting that combining fast context inference with slower, more generalizable feature learning is a fruitful strategy for building more adaptive artificial systems and for understanding how the brain accomplishes this feat.

\section{Setup}

\subsection{Task Structure}

As illustrated in Figure~\ref{fig1}A, our task requires the model to decide whether an image from the $3$D Shapes dataset~\citep{3dshapes18,kim2018disentangling} satisfies a hidden rule. Each rule is defined by one of four latent factors: \emph{floor color} ($10$ values), \emph{wall color} ($10$ values), \emph{object color} ($10$ values), or \emph{object shape} ($4$ values), and each ``rule block'' contains $800$ unique images, half of which align with the rule and thus require a positive response. These blocks appear in random order, with a ``sequence'' comprising $12$ blocks (one block for each of three chosen values across the four factors).

Within each ``experiment,'' we randomly select $3$ values per factor (giving us the ``seen rules'') and run $10$ such sequences. We evaluate how well WiNN and the control models (introduced below) learn and adapt over $10$ different experiments.

To assess generalization, we create a separate test dataset for each experiment. This includes:
\begin{itemize}
    \item \textbf{Validation Set:} $100$ unique images per rule block, used to measure generalization performance on unseen images of an inferred rule.
    \item \textbf{Context-Inference Set:} An additional $800+100$ images per block for testing WiNN's ability to adapt by \emph{only} updating its context state. This set contains:
    \begin{enumerate}
        \item \textbf{Unseen Rules:} $4$ rules (one per factor) that never appeared during training.
        \item \textbf{Seen Simple Rules:} $4$ single-factor rules previously encountered in training.
        \item \textbf{Compositional Rules:} $4$ new rules formed by combining two seen simple rules (e.g., \emph{floor = red} \textbf{AND} \emph{wall = blue}).
    \end{enumerate}
\end{itemize}

Balancing images for single-factor (simple) rules is straightforward, ensuring an equal number of ``yes'' and ``no'' examples. For compositional rules, we ensure an equal number of images following the compositional rule and images following only one of the component simple rules.

\nopagebreak[4]
\subsection{The Wisconsin Neural Network (WiNN)}

The Wisconsin Neural Network (WiNN) brings together four main components to achieve flexible, rule-based visual processing: a backbone, a readout, an attention layer, and a context-inference module (Figure~\ref{fig1}B). We refer to this system as ``WiNN'' because it aims to capture core computations underlying flexible visual problem-solving behaviors, much like those tested in the classic Wisconsin Card Sorting Test.

\paragraph{(1) Backbone.}
We employ a pretrained four-layer convolutional neural network (CNN) to process $64 \times 64$\,px input images. This CNN, trained on the MiniEcoset dataset~\citep{thorat2023characterising}, roughly approximates ventral visual stream computations~\citep{yamins2016using}. Evidence suggests that visual features in the ventral stream are largely acquired during development and change minimally later~\citep{gilbert2012adult,lu2022current,kietzmann2016extensive,cusack2024helpless}. In WiNN, the CNN weights thus remain frozen to preserve learned features, much like stable adult visual processing.

\paragraph{(2) Readout.}
Following the CNN's final layer (AvgPool), we use a linear readout layer that maps visual features to a binary response indicating whether the stimulus conforms to the current rule.

\paragraph{(3) Attention Layer.}
A linear attention mechanism projects the context state multiplicatively onto every neuron in each CNN layer. This mechanism parallels frontoparietal modulation of ventral-stream features~\citep{lindsay2018biological,singercontrasting}, effectively rescaling neuron activations to emphasize features relevant to the inferred rule. Different context vectors yield different patterns of information flow through the CNN, enabling WiNN to produce context-appropriate responses.

\paragraph{(4) Context-Inference Module.}
To allow flexible adaptation across different rules, WiNN maintains a $100$-dimensional context state embedding. Whenever the network produces an incorrect response, this context state is updated iteratively in an effort to reshape the CNN output (via attention) without immediately altering the attention and readout layers. In essence, shifting the context state allows the network to remap its S-R associations in a rapid yet minimally disruptive manner.

\subsubsection{Learning Routine}

During the task, WiNN uses a fast-and-slow learning routine (Figure~\ref{fig1}B, right panel). We present a single stimulus and compute both the network’s accuracy and cross-entropy loss with respect to the correct (binary) response. If an error occurs, we initiate a context inference loop: gradient descent is applied only to the context state for up to $100$ iterations or until the response becomes correct. Conceptually, this rapid adjustment of the context state shifts the network’s attentional focus to produce the correct response for that stimulus, thus achieving a form of S-R remapping. Importantly, the gradient updates do not change any network weights, but are used to change the context state, i.e. the activation pattern of the context embedding. 

After completing (or skipping) the context inference loop, we perform a single gradient descent step on the parameters of the readout and attention layers. The backbone (i.e., the CNN) remains unchanged. This ``slow'' update on the readout and attention parameters minimizes large, rule-specific modifications that might harm generalizability. As a result, WiNN is able to do both, a quick update of the context state to change the information flow through the sensory network to yield correct S-R mapping, and slow learning in the overall readout and attention mechanisms to adapt to the overall dataset statistics. 

We used separate Adam optimizers~\citep{kingma2014adam} for these two updates. The context inference updates had a learning rate of $0.01$ with a weight decay of $0.0001$, while the readout and attention parameter updates had a learning rate of $0.0001$. The context state was initialized to a vector of ones, and all remaining parameters used PyTorch’s default Kaiming uniform initialization~\citep{he2015delving}.

\subsection{Control Models}

One of the main motivations behind the context inference and attention modules in WiNN is to avoid frequent, non-generalizable updates to the CNN backbone. Instead, WiNN treats the CNN as a stable feature extractor that can be selectively modulated by an inferred rule. Without these modules, the CNN itself would have to accommodate S-R remapping. It could do so slowly, risking sluggish adaptation, or more rapidly, risking overfitting to individual stimuli. In contrast, by rapidly adjusting the context state instead of modifying the CNN weights directly, WiNN can potentially generalize from a single example to unseen stimuli: the CNN can reuse its learned features while attention, driven by the context state, shifts to produce the correct rule-based response.

To test this contrast, we considered four control models in which the CNN alone is responsible for S-R adaptation:

\begin{enumerate}
    \item \textbf{CNN-slow (CNN-s).} For each image, the CNN weights receive one update, matching the single update applied to the attention and readout layers in WiNN.
    \item \textbf{CNN-fast (CNN-f).} For each image, the CNN weights receive $101$ updates, mirroring the total of $100$ context updates plus $1$ parameter update in WiNN.
    \item \textbf{CNN-fast-highLR (CNN-fh).} Identical to CNN-fast, but using a learning rate $100$ times higher, matching the context inference rate in WiNN.
    \item \textbf{CNN-fast-lowLR (CNN-fl).} Again identical to CNN-fast, except the learning rate is $100$ times lower, intended to mitigate the risk of catastrophic updates that might arise from multiple updates per image.
\end{enumerate}

\section{Results}

\begin{figure*}
    \centering
    \includegraphics[width=\textwidth]{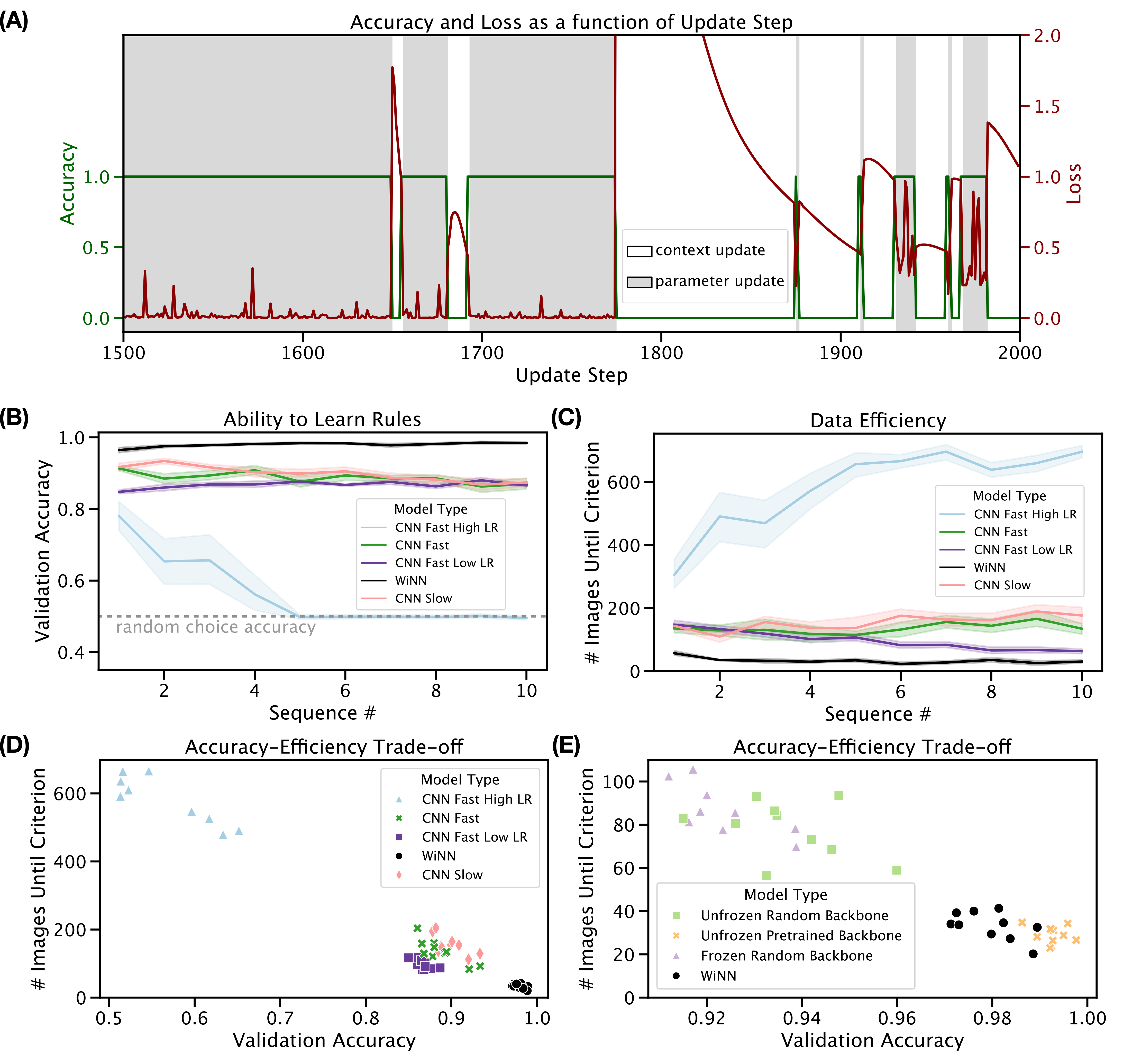}
    \caption{\textbf{WiNN infers rules efficiently.} (A) The learning dynamics of WiNN are shown during a part of an experiment. On the left, towards the end of a rule-block, WiNN is mostly producing correct responses and therefore mostly engaging in attention/readout parameter updates. A rule change elicits a large error, making WiNN engage its context inference loop, minimizing the loss and correcting the response, followed by a single parameter update step. (B) WiNN can infer rules better than the control models - it can better generalize the seen S-R mappings to unseen stimuli. (C) WiNN can infer rules faster than the control models - it needs to see fewer stimuli to be accurate on subsequent stimuli. (D) In sum, efficient rule inference is better in WiNN than in the CNN control models. (E) Unfreezing the backbone still allows efficient rule inference, however, as seen in the Appendix, Figure~\ref{fig:a1} generalization to unseen rules is harder. On the other hand, pretraining the backbone is important for WiNN's efficient rule inference capability.}
    \label{fig:2}
\end{figure*}

\begin{figure*}
    \centering
    \includegraphics[width=\textwidth]{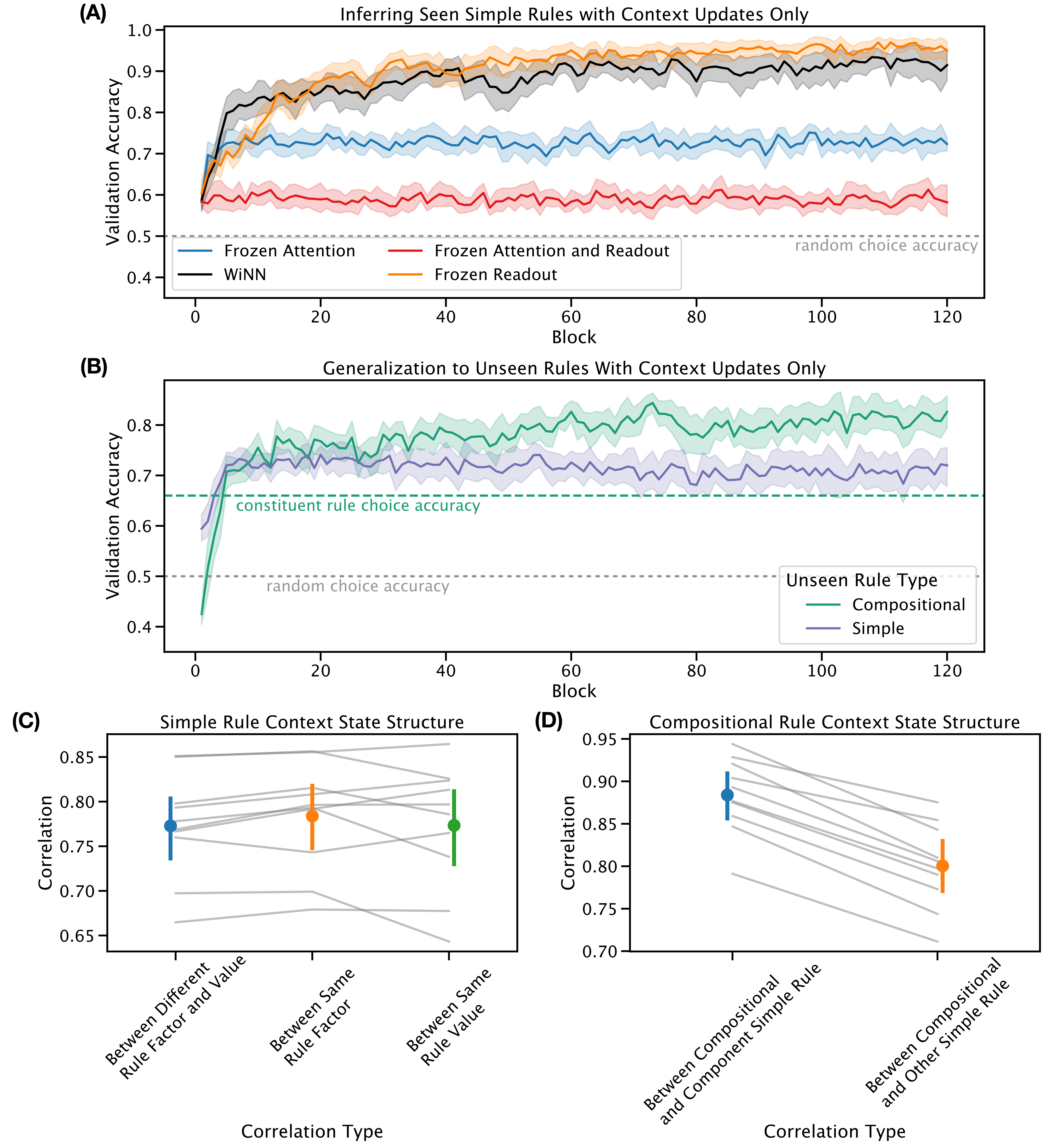}
    \caption{\textbf{WiNN can infer rules solely with context state updates.}
    (A) WiNN can infer seen rules solely with context state updates. This inference becomes better through the experiment. Ablation analyses suggest that the slow attention/readout learning is key to this ability - they ``prepare the ground'' for context-only inference.
    (B) WiNN can also infer unseen simple rules and compositions of seen rules, although not perfectly. This suggests that the learned attention/readout mappings to and from the backbone are general enough to extend to unseen rules.
    (C) The generalization to seen simple rules is not accompanied by an interpretable context state space - neither the latent factors (e.g. wall/shape) nor the color values (e.g. red/blue) are reflected in the context space structure.
    (D) The generalization to compositional rules is accompanied by compositional similarity in the context space - the compositional context state is more similar to the component context states than the non-component context states.}
    \label{fig:3}
\end{figure*}

\subsection{Learning Dynamics of WiNN}

A closer look at WiNN's learning dynamics reveals how it alternates between fast and slow learning during the experiment. As illustrated in Figure~\ref{fig:2}A, whenever WiNN produces a response error (here, during a block transition), it launches a context update loop (white sections) that applies multiple gradient descent steps solely to the context state. This process lowers the cross-entropy loss and corrects the network’s output for that specific stimulus. As a second step, a single gradient descent update is then applied to the attention and readout parameters, leading to a further reduction in loss (gray sections). 

Once a new stimulus arrives that elicits another error, the routine begins anew. If WiNN’s response is already correct upon the new stimulus presentation, the context update loop is skipped and only the parameters of the attention and readout layers receive the usual single update (here, the large gray section on the left, corresponding to the final images of a block). This interplay between rapid context-state updates and slower parameter updates allows WiNN to quickly remap its responses while maintaining general-purpose attention and readout mechanisms.

\subsection{WiNN can infer rules}

In learning S-R mappings, a cognitive system should be able to identify the underlying rule i.e. the factors that allow generalization of the mappings to new stimuli. We tested for this capability by examining whether, after learning a rule, WiNN and the control models could generate correct responses to unseen images. Concretely, at the end of each block in each experiment, we froze the model parameters and measured validation accuracy on a held-out set of images. We then tracked how this accuracy changed over the sequence of blocks (averaged across experiments).

As shown in Figure~\ref{fig:2}B, WiNN maintains a high validation accuracy throughout, indicating that it effectively infers each rule and generalizes beyond the training images. While the CNN-based control models also exhibit reasonable accuracy, they show more pronounced signs of overfitting, with lower overall validation performance compared to WiNN. Notably, the CNN-fh variant (light blue) demonstrates catastrophic overfitting, reaching near-random performance on the validation set. These findings suggest that, although the control models can learn the rules, WiNN’s fast-and-slow learning routine leads to stronger generalization on unseen images.

\subsection{WiNN requires fewer examples to infer rules}

Beyond simply identifying the hidden rule, a cognitively flexible system should be able to do so efficiently (i.e. with fewer examples). We tested for this by measuring how many images per block were required for each model to reach a predefined accuracy threshold, defined as $10$ consecutive images above $90$\% accuracy.

Figure~\ref{fig:2}C shows that WiNN meets this criterion using substantially fewer images than the control models. This confirms WiNN's more efficient rule inference. Consistent with the earlier observations of overfitting, the CNN-fh variant eventually reached the threshold but required significantly more examples than WiNN, despite having the same learning rate for rule-specific updates as WiNN’s context state. 

Together with the validation accuracies reported above, these results suggest that WiNN can both infer new rules and do so using fewer training examples than the control models. This advantage becomes especially clear in Figure~\ref{fig:2}D, which illustrates how WiNN, by the final sequence of the experiment, demonstrates a level of flexibility beyond what simple CNN-based approaches can achieve under similar conditions.

\subsection{The stability gained by freezing the CNN yields the best backbone for contextual inference}

One assumption of WiNN in its current form is that using a pretrained, frozen CNN provides crucial feature stability. To test this assumption we compared WiNN to three variants:
\begin{enumerate}
    \item A randomly-initialized but frozen CNN.
    \item A pretrained but unfrozen CNN, which participated in the slow weight updates.
    \item A randomly-initialized and unfrozen CNN.
\end{enumerate}
We assessed these variants on how efficiently they infer the hidden rule, using the same analysis as in Figure~\ref{fig:2}D.

As shown in Figure~\ref{fig:2}E, including the CNN as part of the slow-update stage slightly improved validation accuracy but did not significantly alter the speed of rule inference. However, supplementary analyses indicated that unfreezing the CNN adversely affected generalization to new rules and compositions of previously learned rules (see Figure~\ref{fig:a1} in the Appendix). Pretraining, however, proved essential: both validation accuracy and inference efficiency dropped when using a randomly-initialized CNN, regardless of whether it was frozen. Overall, these results underscore that freezing a pretrained CNN, consistent with neuroscientific evidence of rather stable adult visual representations, provides the most robust foundation for contextual inference in WiNN.

\subsection{WiNN can infer previously seen rules using only context updates}

A key hypothesis behind decoupling the context state from the attention mechanism is that it creates an ``abstracted'' rule or context space, which WiNN can navigate purely via context updates in response to errors. To examine how much of WiNN’s remapping can be driven solely by adjusting the context state, we performed diagnostic tests at the end of each experimental block. Specifically, we stored the entire network state at the end of each block, and then, for each rule encountered during the experiment, we ran only the context-inference loop (up to $100$ updates each) on a new set of $800$ images, followed by measuring WiNN’s validation accuracy to quantify rule inference.

Figure~\ref{fig:3}A shows that, as the experiment progressed, WiNN’s \emph{context-only} inference achieved near-perfect accuracy in later blocks. Two main explanations could account for this improvement:
\begin{enumerate}
    \item The slow learning of attention and readout parameters is crucial for enabling the context state to drive S-R remapping effectively.
    \item The context state itself benefits from multiple updates, gradually discovering a subspace that allows robust remapping independently of the slow parameter adjustments.
\end{enumerate}

To distinguish between these possibilities, we performed ablation experiments by freezing either the readout, the attention, or both modules in an untrained state and then re-evaluating the context-only inference. As seen in Figure~\ref{fig:3}A:
\begin{itemize}
    \item \textbf{Freezing only the readout.} This did not significantly impair context-only inference, suggesting that a random readout can still be compensated by adaptive attention parameters and context state.
    \item \textbf{Freezing only the attention.} This diminished context-only inference, implying that a fixed attention module cannot be fully compensated by the readout layer and context state alone.
    \item \textbf{Freezing both readout and attention.} This severely reduced validation accuracy, showing that context state alone cannot adapt to a completely random backbone.
\end{itemize}

Together, these findings indicate that the slow learning of attention and readout parameters is indeed essential to ``prepare the ground'' for context-only inference. Once this foundation is established, WiNN can rely increasingly on the context state for rule inference, ultimately achieving near-perfect accuracy without requiring further parameter changes.

\subsection{WiNN can infer unseen rules using only context updates}

The previous section showed that the attention and readout parameters provide a foundation for purely context-based inference. What remains to be tested is, whether these parameters are specifically tuned to the seen rules, or whether they yield generalizable mappings that enable the context state to adapt to unseen or compositional rules. To investigate this, at the end of each block, we again froze WiNN’s parameters and tested its ability to infer unseen simple rules and unseen compositional rules (i.e., conjunctions of familiar rules). For each of these new rule-blocks (each containing $800$ images) only the context state was updated and no other parameters changed. We report the validation accuracies.

Figure~\ref{fig:3}B shows that context-only inference generalizes to unseen rules and compositional rules, although not as effectively as for previously learned rules. Interestingly, while accuracy for unseen simple rules stabilized (with a slight reduction after the first few sequences), performance on conjunction rules improved steadily over time. These findings suggest that the attention and readout layers learn to effectively, and in a generalizable manner, steer the CNN’s information flow, thus enabling the context state to flexibly adapt to novel tasks with no parameter changes.

\subsection{The learned context space of WiNN}

We next investigated whether WiNN’s ability to generalize to unseen simple rules and compositional rules reflects an interpretable organization in its context space. Specifically, we asked whether context states for unseen rules are more similar to context states of previously seen rules that they share latent factors (e.g. wall or object) or color values with. For the compositional rules, we tested whether their inferred context states are more similar to those of their component simple rules than to those of unrelated simple rules.

At the end of each experiment, we froze WiNN’s parameters and ran context inference for the unseen simple rules, the seen simple rules, and their compositions. The resulting context states were stored for later analyses. We excluded shape-related rules when analyzing color, to ensure that any unseen simple rule could share either a latent factor or a specific color value with some seen rules. As a measure of context-state similarity, we used Pearson correlations.

Figure~\ref{fig:3}C shows the average correlation (across $10$ experiments) between unseen simple rules and three categories of seen rules: those matching latent factors, those matching specific color values, and those matching neither. A one-way repeated-measures ANOVA ($F_{2,18} = 1.09$, $p = 0.36$) revealed no significant difference among the three categories. In contrast, Figure~\ref{fig:3}D shows that for conjunction rules, the correlations with their component simple rules were significantly higher than correlations with non-component rules (paired $t$-test, $t_9 = 13.31$, $p = 3.10^{-7}$). Thus, while we see no strong evidence for factor- or color-based structure in the simple-rule context states, compositional rules do display meaningful compositional signatures in WiNN’s context space.

These results suggest that WiNN’s attention/readout parameters and CNN backbone facilitate a form of compositional representation, yet the simple-rule context space may not reflect the underlying latent factors in a straightforward way. Future work might explore whether alternative network architectures or training objectives could yield more disentangled feature representations, thereby enhancing such context-based generalization. 

\section{Discussion}

Potential mechanisms that enable flexible behavior to adapt to changing contexts have been subject to extensive study in both brain science and artificial intelligence. In this paper, we introduced the Wisconsin Neural Network (WiNN), which infers hidden rules and rapidly remaps its S-R behavior by modulating the information flow through a pre-trained feature extractor (CNN) via context-driven attention. Over the course of training, WiNN’s attention and readout parameters learn increasingly general mappings to and from the CNN’s features, ultimately enabling the model to infer the current rule solely by adjusting its context state. Moreover, WiNN’s ability to extend these context-based inferences to unseen and compositional rules illustrates a promising degree of cognitive flexibility. 

\paragraph{Relation to neural accounts of cognitive flexibility.}
WiNN’s fast-and-slow learning algorithm builds on the approach of \citet{hummosthalamus}, which draws inspiration from models of thalamocortical and basal ganglia interactions that underlie context-switching in response to environmental feedback~\citep{wang2021thalamocortical,hummos2022thalamic,zheng2024rapid}. One notable omission in our current framework is a memory mechanism. In the brain, short-term (working) memory integrates information from multiple observations before updating the rule representation, while long-term (episodic) memory can expedite context inference by retrieving previously stored exemplars~\citep{lu2024episodic}. Another aspect of cognitive flexibility, possibly specialized in humans, is the reliance on linguistic or symbolic reasoning~\citep{cole2013rapid}, potentially allowing zero-shot adaptation to new rules~\citep{riveland2022neural}. Incorporating these memory and language-based processes into WiNN could bring it closer to fully capturing the richness of biological cognitive flexibility.

\paragraph{Relation to continual learning research.}
WiNN operates in an online (single-example) learning regime~\citep{van2024continual}. This continual learning setting poses a strong challenge for typical machine learning algorithms. These algorithms rely on batches of randomly sampled data to estimate generalizable gradient updates, without which they suffer catastrophic forgetting. While continual learning techniques such as EWC~\citep{kirkpatrick2017overcoming} and GEM~\citep{lopez2017gradient} mitigate catastrophic forgetting, they rely on task labels to drive their consolidation strategies. WiNN, by contrast, infers changes in the underlying rule from feedback alone (which does not provide rule information explicitly). This aligns with contemporary efforts in task-free continual learning~\citep{collins2019automatically,aljundi2019task,hadsell2020embracing}, which seek to dynamically detect task boundaries to drive consolidation strategies preventing catastrophic forgetting. Therefore, advances to WiNN-like systems, for example from neuroscientific consideration mentioned above, holds promise for continual learning research. 

\paragraph{Comparisons to other fast adaptation methods.}
The idea of using specialized mechanisms for rapid adaptation has also been explored in model-based/meta-learning approaches (e.g., MAML~\citep{finn2017model}, Reptile~\citep{nichol2018first}). These methods learn to quickly retrain or finetune on new tasks with minimal data. WiNN shares the motivation of fast adaptation but differs in its emphasis on context-based modulation of a backbone with frozen weights, coupled with selective updates to attention and readout parameters. The result is a system that can pivot to new tasks or rules using only context inference, bypassing the need for large-scale retraining.

\paragraph{On compositionality.}
A striking ability of WiNN is to infer unseen compositions of previously learned rules. Does this ability truly reflect compositional processes akin to those in the brain? In our task, compositional rules require detecting two bound features simultaneously (e.g., wall = blue \& object = cube), which in WiNN is achieved by modulating all relevant features in parallel. However, according to the influential feature-integration theory of attention~\citep{treisman1980feature}, humans cannot attend to multiple distinct features at once without some form of serial processing. We can indeed multiplex spatial attention with feature attention (e.g. look for blue in the central-right part of the image corresponding to a wall) but detecting a blue wall and a cubical object requires distinct attentional foci and serial processing. 

By contrast, WiNN currently lacks mechanisms for explicit serial or hierarchical attention shifts; instead, its compositional inference arises because the context state for the combined rule overlaps with that of its component simple rules. This yields partial generalization, as evidenced by the lower accuracy for compositional rules compared to familiar single-factor rules (Figure~\ref{fig:3}). Introducing a serial processing capability or a hierarchical rule mechanism could further enhance WiNN’s cognitive flexibility. Such an extension may align more closely with neural evidence of rule hierarchy in the frontal cortex~\citep{badre2009rostro} and provide a foundation for planning and other complex behaviors reliant on nested or sequential rule representations~\citep{botvinick2012hierarchical}.

\paragraph{Conclusion.}
In summary, our in-depth analyses of WiNN indicate that it exhibits \emph{sparks} of cognitive flexibility by balancing rapid context inference with slower, generalizable attention/readout learning. This balance enables WiNN to adapt to changing or unfamiliar rules quickly, while mitigating catastrophic forgetting and preserving stable feature representations. Building on these foundations, by adding richer memory systems, language-based context inference, or hierarchical/serial attention, could advance WiNN’s capabilities and deepen its relevance both to neuroscience and the broader field of continual learning.

\section{Acknowledgments}

The project was financed by the funds of the research training group “Computational Cognition” (GRK2340) provided by the Deutsche Forschungsgemeinschaft (DFG), Germany and the European Union (ERC, TIME, Project $101039524$).

\bibliographystyle{ccn_style}

\bibliography{ccn_style}
\section*{Appendix}

\begin{figure*}
    \centering
    \includegraphics[width=\textwidth]{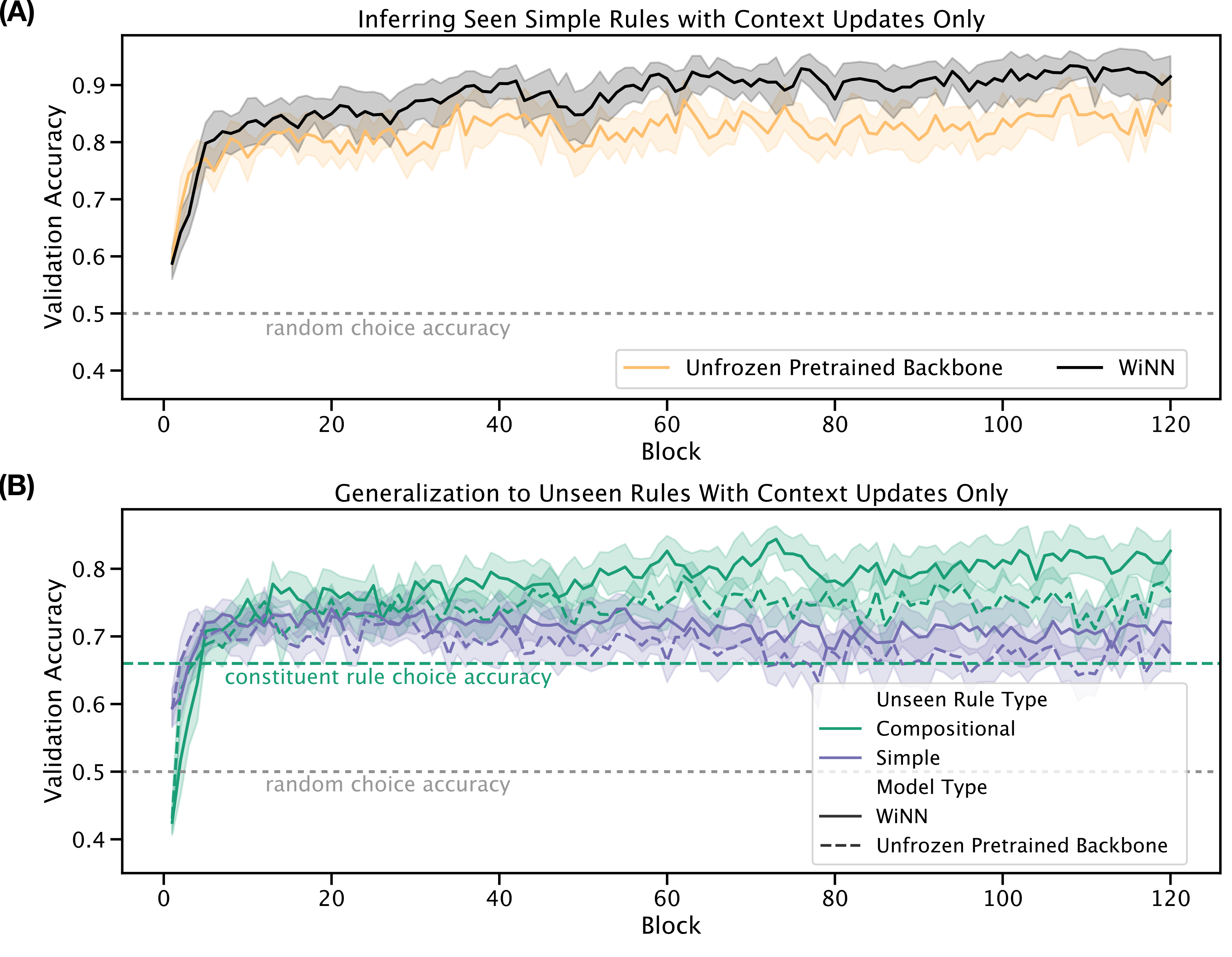}
    \caption{With the same analysis as in Figure~\ref{fig:3}, we observe that the context-only inference capability is worse in the WiNN-variant where the pretrained backbone is unfrozen than in WiNN.}
    \label{fig:a1}
\end{figure*}

\subsection*{WiNN with an unfrozen pretrained backbone has worse generalization to unseen rules}
In the experiments presented in the main text the CNN backbone (sensory network) was frozen and only the attention and readout weights are learned. This reflects the observation that learning in sensory regions (such as the ventral visual stream) is very slow \citep{gilbert2012adult,lu2022current,kietzmann2016extensive,cusack2024helpless}. As an additional control we repeated our experiments starting from the same pretrained CNN but with its weights trainable throughout the experiment.
While this modified WiNN can still perform well on rules it has previously been trained on through context inference, it does so with lowered performance compared to WiNN with a frozen sensory network (Figure~\ref{fig:a1}A). This suggests that WiNN's context inference mechanism benefits from a stable set of features that can be attended to.

Additionally, unfreezing the CNN backbone leads to degraded generalization to both compositional rules created from simple rules the network is exposed to during training (Figure~\ref{fig:a1}B, green lines) as well as to unseen simple rules (Figure~\ref{fig:a1}B, purple lines). 
This suggests that the space of context states that this model has learned is less general and compositional than that of a model with constant feature extractors.
One possible explanation for this is that unfreezing the backbone allows the model to 'overfit' to rules it is trained on, learning features specialized to tasks it is trained on rather than more general features that can be adapted to a wider range of tasks through context-inference and resulting attentional modulation.

\end{document}